\documentstyle[12pt,moriond]{article}
\input{psfig.tex}
\begin{document}
\def\cd{{\it cd}}
\def\HII{H{\sc ii}}
\def\Msun{\, {\rm M}_{\odot}}
\def\Zsun{Z$_{\odot}$ }
\heading{CHEMICAL ABUNDANCE RATIOS \\
AND GAS-PHASE MIXING \\
IN DWARF GALAXY EVOLUTION}

\author{G. Hensler, A. Rieschick} {Institut f\"ur Theoretische Physik 
und Astrophysik,} {Universit\"at Kiel, 
D--24098 Kiel, Germany; email: hensler@astrophysik.uni-kiel.de} 

\begin{moriondabstract}
Because of their low gravitational energies dwarf galaxies
are greatly exposed to energetical influences by the interstellar
medium, like e.g.\ stellar radiation, winds or explosions, 
or by their environment. While the metallicity depletion 
in dwarf galaxies can be explained in general by 
supernova-driven galactic winds, the 
reason for their low N/O ratios at low O abundance is not yet 
completely understood. Stellar yields enrich the different 
gas phases with elements that are characteristic for the stellar 
progenitors. Phase transitions are necessary for a mixing of
elements, but depend sensitively on the thermal and dynamical 
state of the interstellar medium. Models of chemical evolution
start usually with a high N/O ratio at low O abundance according 
to a metal enrichment by ancient stellar populations with 
common yields, but can only reproduce the N/O-O peculiarity 
by the application of multiple starbursts. Their galactic winds
are invoked to reduce O selectively. Chemodynamical models of 
dwarf galaxies, however, demonstrate that strong evaporation
of clouds by the hot supernova gas leads to an almost perfect
mixing of the interstellar gas. These models can
successfully account for the observed N/O-O values in a 
self-consistent way without the necessity of starbursts, 
if new stellar yields are taken into account which provide
additional secondary N production from massive stars. 

\end{moriondabstract}

\section{Introduction}

%\subsection{Evolution of Dwarf Galaxies}

Dwarf galaxies (DGs) present a variety of morphological types.
Their structural and chemical properties differ from those of 
giant galaxies. In addition, low-mass galaxies seem to 
form at all cosmological epochs and by different processes.
Dwarf elliptical galaxies (dEs) are an extremely common and 
astrophysically interesting class of galaxy.  Most known dEs 
are found in regions with high galaxy densities, and they are 
the most numerous of all galaxy types in the cores of nearby 
galaxy clusters (see e.g., \cite{WG84, Bin94}). 
Since the bulk of their star formation (SF) occurred in the past 
(see review by \cite{FB94}), dEs thus are frequently 
considered to be ``stellar fossil'' systems. 
Yet many dwarf spheroidals (dSphs) which 
represent the low-mass end of DGs show not only a significant 
intermediate-age stellar population \cite{Hod89,Gre97}, 
but also more recent SF events \cite{Seta94, Heta97} with 
increasing metallicity, indicating that gas was kept in the system.
In galaxies where gas is depleted by astration stellar abundances 
are predicted to be near the solar value. On the other hand,
the moderate-to-low stellar metallicities in dE and the related 
dSph galaxies (about 0.1 of solar or less) suggest that 
extensive gas loss occurred during their evolution by means of
a supernova typeII (SNII) driven galactic winds (\cite{Lar74,DS86}). 

As another DG type, that consists of the same and higher 
gas fraction as giant spiral 
galaxies (gSs), dwarf irregular galaxies (dIrrs) appear 
with a wide range but lower $Z$ as gSs. 
Again this implies that the metal-enriched 
gas from SNeII was lost from the galaxy. dIrrs show a large 
variety of SF rates from moderate to extraordinarily high values
in starbursts (SBs) and are forming very 
compact star clusters which are embedded in at least one 
intermediate-age to old underlying stellar population. 
Because of their low binding energy SBDGs are characterized by
superwinds \cite{Meta95} or large expanding X-ray plumes
which are driven by SNeII. Their existence
demonstrates the occurrence of metal loss by means of 
large-scale galactic winds and allows to study this phenomenon 
still today. 

The faint blue galaxies at medium redshifts have experienced 
strong SBs and show signatures of present \HII\ galaxies 
\cite{Kota95} what might represent the formation epoch
of the older stellar populations in SBDGs. 
Nevertheless, very metal-poor dIrrs like e.g.\ I~Zw~18 
\cite{Keta95} or SBS~0335-52 \cite{Teta97} seem to form 
their first generation of stars today. In addition, a strange
but perhaps common mode of DG formation exists in tidal tails 
of merging galaxies \cite{Meta92,DM98}. 

%\subsection{Element Abundances and the Chemical Evolution of Dwarf Irregulars}

While the low metal content in DGs can be attributed to
this metal-enhanced mass loss, another puzzling fact needs 
explanation: Why do DGs, though with O abundances below 1/10 
solar, show also low N/O ratios of almost 0.7 dex smaller than 
in gSs with a large scatter and no significant correlation with 
O/H (see fig.1)?

\begin{figure}
% psfile=#1 vsize=#2 angle=#3 hscale=#4 vscale=#5 hoffset=#6 voffset=#7
\psfig{figure=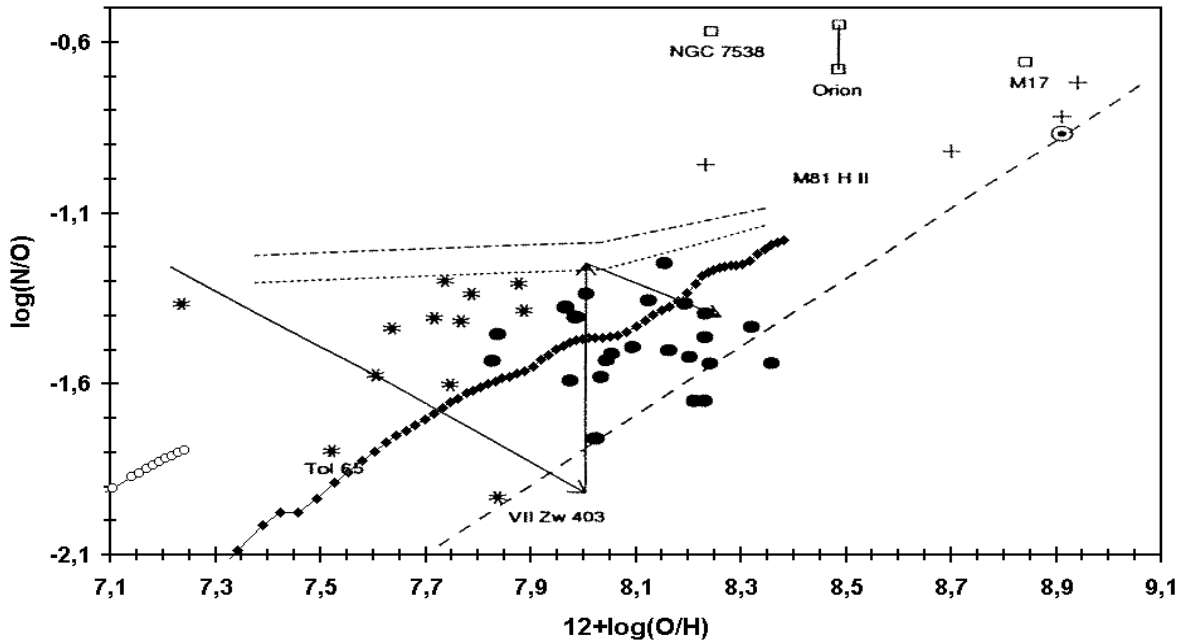,height=8.5cm,width=15.5cm,angle=-90}
\caption[]{Evolutionary tracks of 2d chemodynamical models for 10$^9 \Msun$
         galaxies with (long curve with diamonds) and without DM halos 
         (open circles) in comparison with 
         N/O vs.\ O/H measurements of irregular galaxies (stars and full
	 dots) and two chemical evolutionary models by 
	 \cite{MT85} (upper lines) 
         and a simple model track (arrows; see \cite{Gar90}). }
\end{figure}

While C and N are mainly contributed by planetary nebulae (PNe) from
intermediate-mass stars (IMS) to the warm cloudy gas phase (CM)
of the the interstellar medium (ISM), 
O and Fe are the dominant tracers of SNII and SNIa ejecta, 
respectively, and are therefore initially encorporated in 
the hottest phase of the ISM, the intercloud medium (ICM). 
Various authors have successfully reached the observed 
N/O-O regime of DGs in different chemical models. 
\cite{MT85} and \cite{Meta94} took galactic 
winds into account, which are driven by numerous SB episodes, 
in order to yield the necessary reduction of O. Similarly,
\cite{Gar90} and \cite{Pil92} report models that follow 
the assumption of abundance self-enrichment within the observed 
\HII\ regions \cite{KS86}. All these models differentiate the 
O and N release in accordance to their progenitor lifetimes, 
by this, leading to zick-zack evolutionary tracks in the 
log(N/O)-log(O/H) diagram (\cite{Pil92} and fig.1).
Parameters like wind efficiency, IMF, and gas mass fraction make 
sufficiently good model approachs to the observations possible. 
The yields of these models are in principle based on almost 
the same stellar evolution calculations 
(N by \cite{RV81} and O by \cite{WW86} or by \cite{Mae92} 
and \cite{MM89}). Because of the assumption of an already existing 
element enrichment by ancient stellar populations, the
chemical evolutionary models start 
already from high N/O values at low O abundance.

Since N and O are polluting the different phases of the ISM separately,
their simultaneous existence in \HII\ regions, which represent the ionized
CM, can only result from phase transitions between CM and ICM. 
In models where these processes are taken into account, the 
observed N/O ratio should, therefore, permit qualitative studies 
about both the mixing direction and efficiency. Additionally, its radial 
distribution in a DG provides an insight into dynamical effects of 
the ISM. The assumption of an individual self-enrichment of \HII\ 
regions during their ionization-caused observability \cite{Pil92}
would plausibly involve both, abundance variations during their visibility
epoch and, by this, differences between \HII\ regions within
the same dIrr. In contrast, \cite{Kob98} reports the non-detection
of any sizable O, N, and He anomalies from \HII\ regions in the 
vicinity of young star clusters in SBDGs 
with one exception, NGC~5253, which reveals a central N overabundance. 
In NGC~1569, that has recently formed two super-star clusters, 
\cite{Kob98} finds constant O and N/O values over a radial extent 
of more than 
400 pc with a scatter of N/O by only 0.2 dex, while self-enrichment
tracks extent over one dex in N/O \cite{Pil92} and dispersal
distances are much less. The same fact holds in I~Zw~18 
(Izotov, this conference).

\section{Chemodynamical Models of Dwarf Irregulars}

Not all observed dIrrs that cover the same N/O-O regime 
have passed or experience at present strong SBs, but form
stars at an almost moderate and continuous rate. 
Empirical studies and theoretical investigations 
of systems at low potential energies have decovered that 
their ISM should be balanced by counteracting 
processes like heating and cooling, turbulence and dissipation 
\cite{BH89}. The SF is self-regulated under various conditions 
(see e.g.\ \cite{FC83,Keta95,Keta98}). 
If the evolution of galaxies is sensitive to the energetical
impact of different processes, an adequate treatment of the 
dynamics of stellar and gaseous components and of their mutual 
energetical and materialistic interactions is essential. 
This modelling of galactic evolution is properly performed by the 
{\bf chemodynamical} (\cd ) prescription (for its formulation see 
e.g.\ \cite{TBH92,SHT97}).
The evolution of DGs can proceed in self-regulated ways
both, globally by large-scale flows of unbound gas but also 
locally in the SF sites. \cd\ models of non-rotating DG systems 
achieve SBs from the initial collapse and after mass-loss 
induced expansion in the recollapse phase of its bound 
fraction of the ISM \cite{Heta93,Heta98}. Particular models
are evolving also by oscillatory SF episodes. 
The stellar populations represent dSphs as well as more massive 
{\it blue compact} DGs and reveal a central
concentration of the recent SF region embedded in older elliptical 
populations. External effects \cite{Vil95} like extended 
dark matter (DM) halos, 
the intergalactic gas pressure, etc., could cause further 
morphological differences of DGs e.g.\ by regulating the outflow
and allowing for a recollapse that fuels subsequent SF.

For rotating dIrrs 2d \cd\ models are necessary. 
A gaseous protogalactic cloud starts with or
without a DM halo \cite{Bur95} and with a Plummer-Kuzmin-type 
gas distribution. In the following we wish to discuss only
10$^9 \Msun$ DG models which start with a 2 kpc 
baryonic density scalelength. 
As in former 2d \cd\ models for massive gSs \cite{SHT97}
we have, at first, divided the yields of N and O between IMS and 
massive stars (HMS) according to \cite{MM89,Seta92,Seta93} and to
\cite{WW86}, respectively. Surprisingly, because the \cd\ prescription 
also differentiates the N and O pollution to different gas phases 
and distinguishs between their progenitors' lifetimes, but 
in agreement with the above-mentioned purely chemical models of 
different authors, our \cd\ models reach very rapidly a  
too high log(N/O) ratio between -1.3 to -1.4 at low O 
abundance of almost 7.2 in 12+log(O/H) (see fig.1). Moreover,
without any burst behaviour the further evolutionary track 
surpasses the regime of the observed N/O-O values for dIrrs 
\cite{RH98}. In contrast to the expectation that only the 
ICM is expelled from the dIrrs carrying away the O yield, 
a substatial mixing of CM and ICM determines the 
chemical evolution. Since the hot material cannot completely 
condense onto the clouds, nearly total evaporation of 
the CM in the vicinity of the SF and SNII explosion sites 
must be responsible. This is displayed in the left panel
of fig.2. 
Reasonably, in this case the N mixes perfectly  with O in 
the ICM to an almost constant abundance ratio and can spread 
over larger distances within the DG (right panel of fig.2). 
Due to cooling and dynamical shocks the ICM forms new 
condensations of CM on all galactic scales. The right 
panel of fig.2 seems to reveal a strong outflow, however, 
the greayscale reflects the pure N abundance only, 
but not the absolute N density, which is low and would not 
represent a perceivible  mass loss \cite{RH98}.
But even with a galactic wind the N/O ratio would remain constant
because both elements get lost by the ICM. This means that
the dIrr abundance problem cannot be solved by the selective 
expulsion of O.

 \begin{figure}
% psfile=#1 vsize=#2 angle=#3 hscale=#4 vscale=#5 hoffset=#6 voffset=#7
\psfig{figure=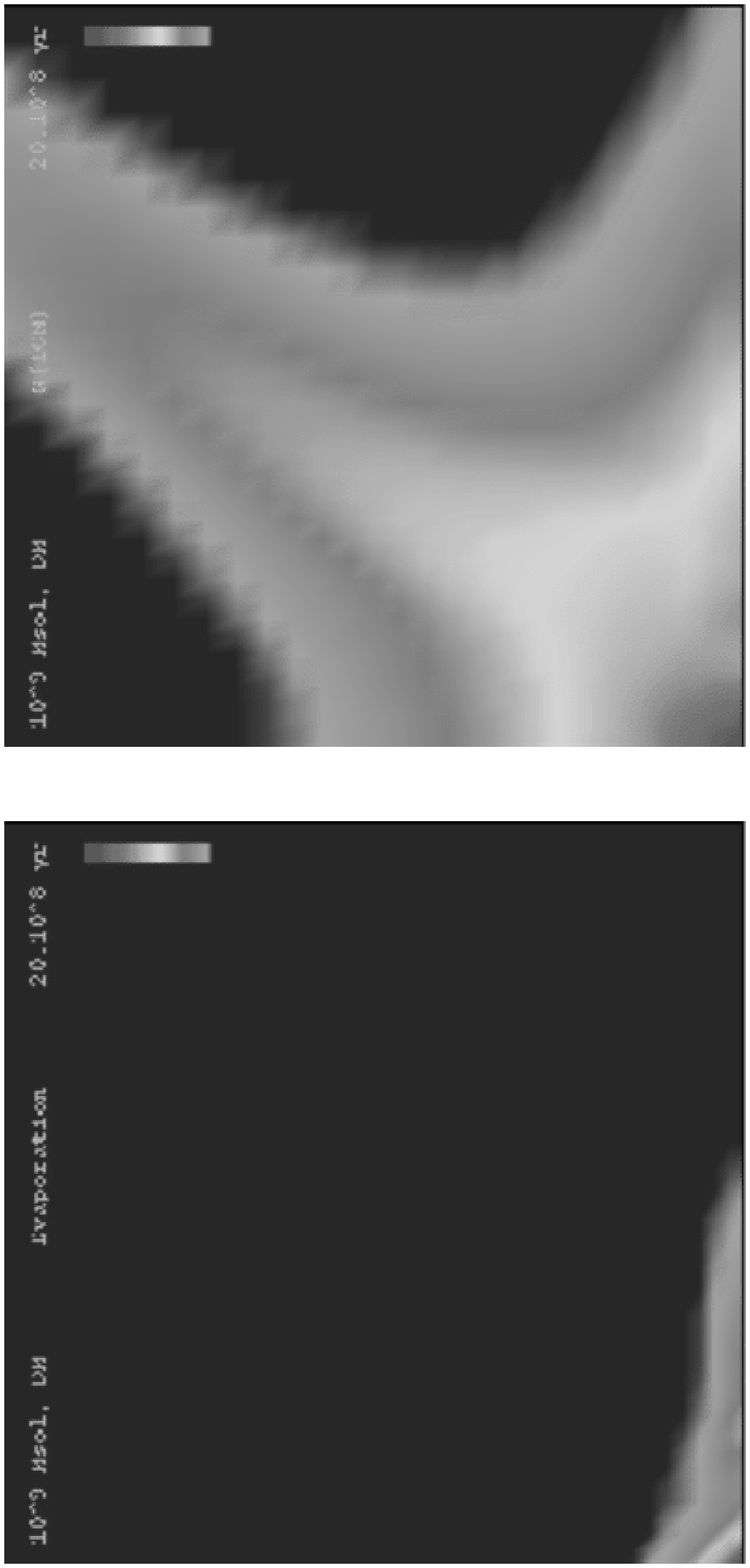,height=8.0cm,width=15.5cm,angle=-90}
\caption[]{Physical condidions of a 2d chemodynamical 10$^9 \Msun$
         dwarf galaxy model with a DM halo after 2 Gyr. 
         The size of the figures is 6.8 kpc. \\
         left: grayscale of the evaporation rate of warm 
	 interstellar clouds by the hot intercloud medium (ICM); 
	 the maximum lies at the lower left corner. \\
	 right: greyscale of N abundance in the ICM; the maximum 
         lies at the lower left corner. \\ }
\end{figure}

Recent stellar evolution models by \cite{WW95}, however, 
provide a secondary N production also by HMS.
Their new yields combined with the most recent models 
for IMS by \cite{vdHG97} have been successfully fitted by 
\cite{Sam98} to the abundance distributions in our \cd\
Milky Way model \cite{SHT97}. 
In contrast to the cited chemical and first 2d \cd\ models of 
DGs which reach too large N/O ratios or even begin in that range,
both our 10$^9 \Msun$ models with new yield 
prescriptions commence at very low N/O and O values 
due to the delayed N release by PNe and the lower 
N production in massive stars at low metallicities.
The N/O-O track of the model with DM halo then rises rapidly,
both N/O and O, and reaches log(N/O) of -1.8 after 1 Gyr 
and -1.6 after 2 Gyrs, respectively (see fig.1). 
These values are only somewhat smaller than the numerical 
yield ratios and are based on an almost perfect 
mixing of CM and ICM. Due to the large-scale hot gas streaming
and subsequent condensation the abundances observed in 
\HII\ regions show a constant value \cite{Kob98}. 
With the same differential mixing processes the 
observed C/O tendency of DGs \cite{Geta95} could also be 
explained \cite{RH98}.

\section{Conclusion}

Here we have shown that in \cd\ models of dIrrs a strong 
mixing of the gas phases prevents a selective element depletion. 
The self-consistent \cd\ models 
evolve with local variations but globally moderate SF 
activity \cite{RH98} and therefore typically for a large
fraction of dIrrs.
The observed N and O abundances in the \HII\ regions
can be strikingly attained, if N is
produced as a secondary element also in HMS. Their
evolutionary tracks of log(N/O)-log(O/H) rise from low values,
reach the regime of observations after almost 3 Gyr and
cover the region for another 3 Gyr depending on the DM halo
contribution. Only for a much older stellar populations 
the existing ISM could be more metal-enriched, so
that it must be partly rejuvenated by infall of 
intergalactic gas or from a bound gas reservoir, both with almost 
pristine abundances. 

The results change, if condensation would dominate
the phase transitions and parts of the ICM 
(and therefore of the O content) would be adapted by the CM. 
Reasonably, this would lead to higher N/O ratios 
and, in addition, to inhomogeneous N/O distributions.
Strong operation of condensation in the SF body of dIrrs
can therefore be excluded but works preferably at higher 
gas densities. 
Because larger gravitational potentials and resulting mass 
densities in more massive galaxies lead to a faster evolution
by a more intense cooling of the ICM and significantly higher condensation 
and SF rates, the N/O increase is stronger, so that discrepancies between
the different yield assumptions (with and without secondary N
production in HMS) cannot be noticed in \cd\ models of
gSs like the Milky Way \cite{SHT97,Sam98}. \\

{\small
Acknowledgments: We are grateful to J.\ K\"oppen, M.\ Samland 
and C.\ Theis for discussions. 
This work is supported by the {\it Deutsche 
Forschungsgemeinschaft} (DFG) under grant no.\ He 1487/5-3 (A.R.).

% (Literaturverzeichnis)

}
\vfill
\end{document}